\documentclass[letterpaper]{article} 
\usepackage{aaai2026}  
\usepackage{times}  
\usepackage{helvet}  
\usepackage{courier}  
\usepackage[hyphens]{url}  
\usepackage{graphicx} 
\usepackage{natbib}  
\usepackage{caption} 
\usepackage[utf8]{inputenc}         
\usepackage{url}                    

\usepackage{booktabs}               
\usepackage{amsfonts}               
\usepackage{nicefrac}               
\usepackage{microtype}              
\usepackage{xcolor}                 

\usepackage{amsmath}
\usepackage{amsthm}
\usepackage{amssymb}
\usepackage{enumitem}
\usepackage{environ}
\usepackage{adjustbox}
\usepackage{subcaption}
\usepackage{anyfontsize}
\usepackage{accents}
\usepackage{colortbl}
\usepackage[flushleft]{threeparttable}
\usepackage{multirow}
\usepackage{siunitx}
\usepackage[algo2e,ruled,noend,noline]{algorithm2e}
\SetKwComment{Comment}{/* }{ */}
\SetKwInput{KwData}{Given}
\LinesNumbered

\usepackage{tikz}
\usetikzlibrary{backgrounds}
\usetikzlibrary{intersections}
\usetikzlibrary{trees}
\usetikzlibrary{fit}
\usetikzlibrary{decorations.pathreplacing,calligraphy}
\usepackage{pgfplots}
\pgfplotsset{compat=newest}
\pgfplotsset{every axis legend/.append style={legend cell align=left}}
\usepgfplotslibrary{groupplots,statistics,patchplots,fillbetween,ternary}

\usepackage{cleveref}

\makeatletter
\newsavebox{\measure@tikzpicture}
\NewEnviron{scaletikzpicturetowidth}[1]{%
  \def\tikz@width{#1}%
  \def\tikzscale{1}\begin{lrbox}{\measure@tikzpicture}%
  \BODY
  \end{lrbox}%
  \pgfmathparse{#1/\wd\measure@tikzpicture}%
  \edef\tikzscale{\pgfmathresult}%
  \BODY
}

\newcommand{\argmax}{\operatornamewithlimits{arg\,max}}

\newcommand{\ra}[1]{\renewcommand{\arraystretch}{#1}}
\newcommand{\grayrule}{\arrayrulecolor{black! 30}\midrule\arrayrulecolor{black}}
\newsavebox\CBox

\definecolor{commentgray}{rgb}{0.4, 0.4, 0.4}
\newcommand{\FlushRightComment}[1]{%
  \hfill{\color{commentgray}$\triangleright$ #1\hspace*{-2.0em}}}

\nocopyright 

\title{Efficient Multiagent Planning via Shared Action Suggestions}
\author {
    Dylan M. Asmar,
    Mykel J. Kochenderfer
}
\affiliations {
    Stanford Intelligent Systems Laboratory, Stanford University, Stanford, CA\\
    asmar@stanford.edu, mykel@stanford.edu
}

\begin{document}

\maketitle

\begin{abstract}
Decentralized partially observable Markov decision processes with communication (Dec-POMDP-Com) provide a framework for multiagent decision making under uncertainty, but the NEXP-complete complexity for finite-horizon problems renders solutions intractable in general. While sharing actions and observations can reduce the complexity to PSPACE-complete, we propose an approach that bridges POMDPs and Dec-POMDPs by communicating only suggested joint actions, eliminating the need to share observations while retaining near-centralized performance. Our algorithm estimates joint beliefs using shared actions to prune infeasible beliefs. Each agent maintains possible belief sets for other agents, pruning them based on suggested actions to form an estimated joint belief usable with any centralized policy. This approach requires solving a POMDP for each agent, reducing computational complexity while preserving performance. We demonstrate its effectiveness on several Dec-POMDP benchmarks, showing performance comparable to centralized methods when shared actions enable effective belief pruning. This action-based communication framework offers a natural avenue for integrating human-agent cooperation, opening new directions for scalable multiagent planning under uncertainty, with applications in both autonomous systems and human-agent teams.
\end{abstract}

\begin{links}
    \link{Code}{github.com/sisl/MCAS}
\end{links}

\section{Introduction} \label{sec:intro}

From complex engineering projects to emergency response teams, effective coordination between individuals is vital for success. The ability of humans to work together, communicate intuitively, and adapt to changing conditions has inspired researchers to explore cooperation in autonomous systems \citep{albrecht_autonomous_2018}. However, achieving such seamless collaboration in autonomous teams remains a significant challenge.

For multiagent decision making under uncertainty, where agents act without complete knowledge and outcomes are stochastic, the decentralized partially observable Markov decision process (Dec-POMDP) \citep{berstein_complexity_2002} is a widely used model. Agents must reason about both their environment and other agents' possible actions and beliefs without direct communication. While powerful, Dec-POMDPs are notoriously hard to solve, making them impractical for many real-world problems \citep{oliehoek_concise_2016}.

When agents can communicate, the computational burden can be reduced under certain assumptions \citep{pynadath_communicative_2002,goldman_optimizinginfo_2003}. However, sharing complete information is often impractical, and when communication is not lossless and free, the complexity of solving a finite-horizon Dec-POMDP with communication (Dec-POMDP-Com) remains NEXP-complete \citep{goldman_coop_complexity_2004}.

As autonomous systems become more capable, human-machine collaboration becomes increasingly relevant \citep{Johnson_ai_island_2019,dafoe_open_probs_2020}. While combining human intuition with machine computation could create superior teams, this requires addressing both multiagent coordination complexities and human-machine communication challenges \citep{grosz_collaborative_plans_1996,Crandall_cooperating_2018,Tabrez_mental_model_2020}. Humans naturally communicate through action suggestions like ``let's move there'' without sharing detailed observations or beliefs. For instance, suggesting ``We should eat at restaurant X'' implicitly communicates beliefs about the restaurant being open, suitable, and reasonably priced, thus encapsulating complex reasoning in a simple action proposal.

This type of action-based communication is natural for humans but has not been fully explored for enabling collaboration in autonomous systems or human-agent teams. We propose an approach that narrows the focus of communication to suggested joint actions. Instead of sharing raw observations or beliefs, agents communicate their recommended actions at each step, mirroring how humans often collaborate in complex tasks and using action suggestions to convey important information about their understanding of the situation.

Our proposed method estimates joint beliefs by maintaining sets of reachable beliefs and inferring other agents' beliefs. The key insight is that an action suggestion implies the agent's belief is within a particular subspace of the belief space, allowing us to prune infeasible beliefs. The agent can then more accurately infer the other agents' beliefs, enabling the construction of an estimated joint belief that can be used with a policy assuming centralized execution. This method requires solving $n$ multiagent POMDPs (MPOMDPs) for an $n$ agent problem, online computation of belief updates for all of the beliefs in the belief set, and a joint policy using the centralized assumptions (solving another MPOMDP). We evaluate this approach on several standard Dec-POMDP benchmarks and more complex variations, demonstrating performance similar to fully centralized methods when shared actions enable effective belief pruning.

\section{Related Work} \label{sec:related work}
This work builds upon key areas in decentralized decision making, including communication in Dec-POMDPs, sufficient statistics for planning, and action-based coordination methods. The introduction of communication to Dec-POMDPs has been explored to reduce computational burden and improve coordination. \citet{pynadath_communicative_2002} examined how communication strategies improve multi-agent teamwork, while \citet{goldman_optimizinginfo_2003} investigated optimizing information exchange in these models. The concept of sufficient statistics has played an important role in simplifying Dec-POMDP planning. \citet{oliehoek2013sufficient} introduced probability distributions over joint action-observation histories as sufficient plan-time statistics. \citet{dibangoye_oMDP_2016} recast Dec-POMDPs as continuous state MDPs using occupancy states, enabling POMDP techniques.

This work is also related to research on action-based coordination in multi-agent systems. Previous work has explored the use of suggested actions as a means of communication between agents, treating these suggestions as observations of the environment \citep{asmar_2022}. However, that work assumed that suggested actions were conditioned on the true state of the environment, which becomes less reliable when agents make suggestions based on their beliefs about the state rather than the state itself. A practical example can be found in aircraft collision avoidance systems like TCAS and ACAS X \citep{asmar_2013_ATC408}, which use action advisories (e.g., "do not descend") to restrict other aircraft's actions, effectively coordinating decisions without direct observation sharing by modifying costs of incompatible actions \citep{asmar_thesis_2013}.

This work builds on the idea that communication of suggested joint actions can help reduce computational complexity. By using these suggestions to construct distributions over other agents' beliefs, we provide sufficient statistics for histories while allowing all agents to maintain partial observability. The approach reduces the need for full communication while maintaining coordination efficiency, and provides a natural framework for human-agent teaming where action suggestions are an intuitive mode of communication.

\section{Background} \label{sec:background}

A partially observable Markov decision process (POMDP) models sequential decision making under uncertainty \citep{Smallwood1973TheOC}. In a POMDP $\left( \mathcal{S}, \mathcal{A}, \mathcal{O}, T, O, R, \gamma \right)$, an agent in state $s \in \mathcal{S}$ chooses an action $a \in \mathcal{A}$, transitions to $s^\prime$ based on $T(s, a, s^\prime) = P(s^\prime \mid s, a)$, and receives an observation $o \in \mathcal{O}$ based on $O(s^\prime, a, o)=P(o \mid s^\prime, a)$. The agent receives a reward $R(s,a) \in \mathbb{R}$ with discount factor $\gamma \in [0, 1)$. One method to solve a POMDP is to infer a belief distribution $b \in \mathcal{B}$ over $\mathcal{S}$ and then solve for a policy $\pi$ that maps the belief to an action where $\mathcal{B}$ is the set of beliefs over $\mathcal{S}$ \citep{Kochenderfer2022}. Executing with this type of policy requires maintaining $b$ through updates after each time step.

A Decentralized POMDP (Dec-POMDP) extends the POMDP framework to multiple cooperative agents with tuple $(\mathcal{I}, \mathcal{S}, \{\mathcal{A}^i\}, \{\mathcal{O}^i\}, T, O, R, \gamma)$. Each agent $i \in \mathcal{I}$ selects an action $a^i \in \mathcal{A}^i$ and receives an observation $o^i \in \mathcal{O}^i$. We use superscripts to represent the agent index and bold variables to represent the joint collection across all agents, e.g., $\mathbf{a} = (a^1, \ldots, a^{|\mathcal{I}|})$. The true state is shared by all agents, while the reward, transition, and observation functions are defined over joint actions and observations \citep{oliehoek_concise_2016, Kochenderfer2022}.

In a Dec-POMDP with communication (Dec-POMDP-Com), agents can communicate using messages from alphabets $\{\Sigma^i\}$ with communication cost function $C_{\Sigma}$ \citep{pynadath_communicative_2002, oliehoek_concise_2016}. In both models, agents make decisions based on their individual action-observation histories (and messages in Dec-POMDP-Com). Solving a finite horizon Dec-POMDP or Dec-POMDP-Com is NEXP-complete \citep{berstein_complexity_2002, oliehoek_concise_2016}. If agents can communicate actions and observations perfectly without cost, the model becomes a multiagent POMDP (MPOMDP), which can be solved using POMDP approaches \citep{pynadath_communicative_2002, oliehoek_concise_2016}.

In our approach, we use MPOMDP policies instead of solving the Dec-POMDP directly. Policies for POMDPs and MPOMDPs can be generated offline or computed online during execution. In this work, we integrate our method with policies generated offline and leave the application to online solvers for future work. In particular, we use SARSOP \citep{Kurniawati2008SARSOPEP} to generate the policies and represent the policy as a set of alpha vectors, but our approach is not limited to SARSOP or alpha vectors and can be applied to policies generated by other methods.

\section{Problem Formulation and Notation}
Our work addresses collaborative sequential decision-making under uncertainty within the Dec-POMDP-Com framework, focusing on infinite-horizon problems with discrete state, action, and observation spaces. Although our methods are presented in the context of infinite horizons, they are also applicable to finite-horizon scenarios. We consider a system of $n$ agents, each receiving individual observations and maintaining their own belief over the state space. The individual belief of agent $i$ at time $t$ is designated as $b_t^i \in \mathcal{B}^i$. At each time step, after performing an action and receiving an observation, agents communicate by suggesting an action that is optimal using their policy. Therefore, the message alphabet for each agent equals their action space $\Sigma^i = \mathcal{A}^i$. Messages are assumed to be transmitted and received perfectly, without cost, noise, or loss. We denote the message from agent $i$ to agent $j$ at time $t$ as $\sigma_t^{i,j} \in \Sigma^i$.

Each agent also maintains a set of possible beliefs that the other agents might possess. For estimations maintained by agents, we use a \textit{hat} symbol. A superscript of two indices $i, j$ on an estimation refers to agent $i$'s estimate about agent $j$. For example, $\hat{b}_{t_k}^{i,j}$ represents the $k^\text{th}$ estimated belief agent $i$ has for agent $j$'s belief at time $t$. Using similar notation, $\hat{\mathcal{B}}^{i,j}$ is the set of all estimated beliefs agent $i$ has for agent $j$'s belief. When referring to joint beliefs, we use a tilde, $\tilde{b}$.

We assume each agent has access to surrogate policies for other agents, where the surrogate policy $\hat{\pi}^{i,j}$ is agent $i$'s model of agent $j$'s policy. In our experiments where we compute the policies in a centralized manner offline, $\hat{\pi}^{i,j} = \pi^j$. In a slight abuse of notation, we use subscripts to indicate the time step ($b_t$), counting of the number of variables of a collection (subscript to the time step, $b_{t_k}$), and for indexing actions and observations ($a_\ell, o_m$). For actions and observations, subscripts reference the index within the space, e.g. $a^i_\ell \in \mathcal{A}^i$ is the $\ell^\text{th}$ action in $\mathcal{A}^i$.

\section{Using Action Suggestions}

There are several ways agents can use suggested actions. The simplest option is to ignore the messages and choose actions as if there was no communication, which is equivalent to a Dec-POMDP. Alternatively, agents could designate a leader at each time step and follow the leader's suggested actions, which is sufficient in some environments where one agent's observations provide enough information, as in the Broadcast Channel problem (\cref{sec:results}). Another approach is hierarchical action selection, where agents select actions and communicate following a specific communication order. In this scheme, each agent can select an action with knowledge of the previous messages received for that time step. The order of communication becomes important as agents earlier in the process have to make decisions with less information. This approach is similar to other prioritization schemes \citep{dibangoye2009topological}.

In our approach, we use suggested actions to infer beliefs. In a cooperative scenario, we assume agents act optimally to maximize shared rewards. Therefore, we assume the suggested action is the one that maximizes the expected sum of discounted rewards based on the agent's belief of the environment. Referencing back to the restaurant example from the introduction, we can infer aspects of the friend's belief from their action suggestion by assuming they are acting optimally and want to maximize the happiness of the group. For instance, if a friend suggests a restaurant, we can infer they believe it is open and suitable for the group's preferences. Each action suggestion thus contains information related to the suggester's belief of the environment, which we can use to infer their belief.

\subsection{Inferring the Belief Subspace}\label{sec:inferring}

We can use the suggested action and the fact that the suggested action is the optimal action from the suggester's perspective to infer the possible beliefs the agent could have. For example, if agent $i$ receives a suggested action $a_s$ from agent $j$ using policy $\pi^j$, then we know $b^j \in \mathcal{B}^j_{a_s}$ where $\mathcal{B}^j_{a_s} = \{b \mid \pi^j(b) = a_s, \forall b \in \mathcal{B}^j \}$.

In an alpha vector policy, this would be the subspace of beliefs that are dominated by alpha vectors associated with the suggested action. With a set of alpha vectors $\Gamma$ representing the policy and a suggested action $a_s$
\begin{align}
    \mathcal{B}^j_{a_s} = \{\mathbf{b} \mid  ( \boldsymbol{\alpha}_i - \boldsymbol{\alpha}_j ) \cdot \mathbf{b} \geq 0, \forall \boldsymbol{\alpha}_i \in \Gamma_{a_s}, \forall \boldsymbol{\alpha}_j \in \Gamma \} \label{eq:alpha_sub}
\end{align}%
where $\mathbf{b}$ is a belief vector representing probabilities over states and $\Gamma_{a_s} \subseteq \Gamma$ is the set of alpha vectors corresponding to action $a_s$. 

\subsection{Pruning Beliefs}\label{sec:pruning}
At each time step, agents update their beliefs based on individual observations and actions performed. From agent $i$'s perspective, there are $|\mathcal{O}^j|\prod_{i \neq j} |\mathcal{A}^k|$ possible beliefs reachable from $\hat{b}_t^{i,j}$ for agent $j$. The size of this set grows exponentially in time, reaching $\left( |\mathcal{O}^j|\prod_{i\neq j} |\mathcal{A}^k| \right)^\ell$ after $\ell$ time steps. This exponential growth is one of the primary factors in the NEXP complexity of solving Dec-POMDPs.

To help manage this growth, we can prune infeasible beliefs using the suggested actions. We can rigorously define the belief subspace in which the suggester's belief must lie (\cref{eq:alpha_sub}) and this subspace is an infinite set of beliefs. While we cannot easily construct the subspace, we can test if a belief is within this subspace by evaluating the policy at that belief.

\begin{figure}[tb!]
    \begin{center}
        \begin{scaletikzpicturetowidth}{0.455\textwidth}
            \begin{tikzpicture}[
    grow'=down,
    level 1/.style={level distance=1.5cm, sibling distance=3.7cm},
    level 2/.style={level distance=1.95cm, sibling distance=1.2cm},
    edge from parent/.style={draw, -latex},
    action/.style={rectangle, draw, fill=white, font=\scriptsize, text width=1.3cm, minimum height=0.3cm, align=center},
    belief/.style={circle, draw, fill=white, minimum size=0.5cm, inner sep=0pt, font=\scriptsize, text width=0.5cm, align=center},
    equations/.style={below=1.0cm, font=\tiny, align=center},
    edge_l label/.style={midway, left, font=\tiny},
    edge_m label/.style={midway, left, font=\tiny, inner sep=1pt},
    edge_r label/.style={midway, right, font=\tiny},
    scale=\tikzscale
]
    \node[belief] {$\hat{b}_{t-1}^{1,2}$}
    child { node[action] (a22) {$(a_1^1, a_2^2, a_2^3)$}
        child { node[belief] (b12) {$\hat{b}_{t_{12}}^{1,2}$} 
            edge from parent node[edge_r label] {$o_3^2$} 
        }
        child { node[belief] (b11) {$\hat{b}_{t_{11}}^{1,2}$} 
            edge from parent node[edge_m label] {$o_2^2$} 
            node[equations] {
                $\color{green!55!black}\hat{\pi}^{1,2}(b_{t_{10}}^2) = \boldsymbol{\sigma}_t^{2,1}$ \\ 
                $\hat{\pi}^{1,2}(b_{t_{11}}^2) \color{red!90!black}\neq \color{black} \boldsymbol{\sigma}_t^{2,1}$ \\
                $\hat{\pi}^{1,2}(b_{t_{12}}^2) \color{red!90!black}\neq \color{black} \boldsymbol{\sigma}_t^{2,1}$
            }
        }
        child { node[belief] (b10) {$\color{green!55!black}\hat{b}_{t_{10}}^{1,2}$} 
            edge from parent node[edge_l label] (og10) {$\color{green!55!black}o_1^2$}
        }
    }
    child { node[action] (a21) {$(a_1^1, a_2^2, a_1^3)$}
        child { node[belief] (b9) {$\hat{b}_{t_9}^{1,2}$} 
            edge from parent node[edge_r label] {$o_3^2$}
        }
        child { node[belief] (b8) {$\hat{b}_{t_8}^{1,2}$} 
            edge from parent node[edge_m label] {$o_2^2$} 
            node[equations] {
                $\hat{\pi}^{1,2}(b_{t_{7}}^2) \color{red!90!black}\neq \color{black} \boldsymbol{\sigma}_t^{2,1}$ \\ 
                $\hat{\pi}^{1,2}(b_{t_{8}}^2) \color{red!90!black}\neq \color{black} \boldsymbol{\sigma}_t^{2,1}$ \\
                $\hat{\pi}^{1,2}(b_{t_{9}}^2) \color{red!90!black}\neq \color{black} \boldsymbol{\sigma}_t^{2,1}$
            }
        }
        child { node[belief] (b7) {$\hat{b}_{t_7}^{1,2}$} 
            edge from parent node[edge_l label] {$o_1^2$} 
        }
    }
    child { node[action] (a12) {$(a_1^1, a_1^2, a_2^3)$}
        child { node[belief] (b6) {$\hat{b}_{t_6}^{1,2}$} 
            edge from parent node[edge_r label] {$o_3^2$} 
        }
        child { node[belief] (b5) {$\color{green!55!black}\hat{b}_{t_5}^{1,2}$} 
            edge from parent node[edge_m label] (og5) {$\color{green!55!black}o_2^2$} 
            node[equations] {
                $\hat{\pi}^{1,2}(b_{t_{4}}^2) \color{red!90!black}\neq \color{black} \boldsymbol{\sigma}_t^{2,1}$ \\ 
                $\color{green!55!black}\hat{\pi}^{1,2}(b_{t_{5}}^2) = \boldsymbol{\sigma}_t^{2,1}$ \\
                $\hat{\pi}^{1,2}(b_{t_{6}}^2) \color{red!90!black}\neq \color{black} \boldsymbol{\sigma}_t^{2,1}$
            }
        }
        child { node[belief] (b4) {$\hat{b}_{t_4}^{1,2}$} 
            edge from parent node[edge_l label] {$o_1^2$} 
        }
    }
    child { node[action] (a11) {$(a_1^1, a_1^2, a_1^3)$}
        child { node[belief] (b3) {$\hat{b}_{t_3}^{1,2}$} 
            edge from parent node[edge_r label] {$o_3^2$} 
        }
        child { node[belief] (b2) {$\hat{b}_{t_2}^{1,2}$} 
            edge from parent node[edge_m label] {$o_2^2$} 
            node[equations] {
                $\hat{\pi}^{1,2}(b_{t_{1}}^2) \color{red!90!black}\neq \color{black} \boldsymbol{\sigma}_t^{2,1}$ \\ 
                $\hat{\pi}^{1,2}(b_{t_{2}}^2) \color{red!90!black}\neq \color{black} \boldsymbol{\sigma}_t^{2,1}$ \\
                $\hat{\pi}^{1,2}(b_{t_{3}}^2) \color{red!90!black}\neq \color{black} \boldsymbol{\sigma}_t^{2,1}$
            }
        }
        child { node[belief] (b1) {$\hat{b}_{t_1}^{1,2}$} 
            edge from parent node[edge_l label] {$o_1^2$} 
        }
    };

    \begin{scope}
        \node[draw=green!55!black, dashed, rounded corners, fit=(b5) (og5), inner sep=0.05cm, line width=2.5pt] {};
    \end{scope}

    \begin{scope}
        \node[draw=green!55!black, dashed, rounded corners, fit=(b10) (og10), inner sep=0.05cm, line width=2.5pt] {};
    \end{scope}
    
\end{tikzpicture}
        \end{scaletikzpicturetowidth}
    \caption{Example of pruning reachable beliefs that do not align with the received message $\boldsymbol{\sigma}_t^{2,1}$. This example has $n=3$, $|\mathcal{A}^i|=2$, and $|\mathcal{O}^i|=3$. The process is from agent $1$'s perspective, expanding a single belief estimate for agent $2$.}
    \label{fig:branching}
    \end{center}
\end{figure}

Without loss of generality, we will discuss this process from the perspective of agent $i$ maintaining a belief estimate for agent $j$. We start with an initial belief set $\hat{\mathcal{B}}^{i,j}_0 = \{ b_0^j \}$, where in our approach, we assume all agents begin with the same initial belief. After performing an action and receiving a local observation, we expand the beliefs considering all possible actions and observations, resulting in $|\hat{\mathcal{B}}^{i,j}_t| = |\hat{\mathcal{B}}^{i,j}_{t-1}| |\mathcal{O}^j| \prod_{j \neq i} |\mathcal{A}^j|$ at time $t$. We then evaluate each belief with the surrogate policy for agent $j$ and prune the beliefs where the optimal action does not match the received message%
\begin{equation}
\hat{\mathcal{B}}^{i,j}_t \leftarrow \{ b \in \hat{\mathcal{B}}^{i,j}_t \mid \hat{\pi}^{i,j}(b) = \sigma^{j,i} \}.
\end{equation}%
\Cref{fig:branching} illustrates this pruning process in an example with three agents. If we know the actions performed at the last time step, we only need to consider observations for a single joint action, increasing our estimated belief set by a factor of $|\mathcal{O}^j|$ instead of $|\mathcal{O}^j| \prod_{j \neq i} |\mathcal{A}^j|$. This knowledge significantly reduces the size of the reachable belief set. 

After pruning the infeasible beliefs, we can further reduce our set by removing beliefs that are sufficiently close to other beliefs in the set. \citet{Zhang_covering_2012} showed that for any two beliefs $b$ and $b^\prime$, if $||b - b^\prime||_1 \leq \delta$, then $|P(o \mid b, a) - P(o \mid b^\prime, a)| \leq \delta$. Additionally, \citet{lee_approximate_2007} proved that the value function of POMDPs satisfies the Lipschitz condition, i.e., $|V(b) - V(b^\prime)| \leq \frac{||R||_\infty}{1-\gamma}\delta$ if $||b-b^\prime||_1 \leq \delta$ and \citet{wu_adaops_2021} used this bound to combine beliefs in their proposed POMDP algorithm. Building on this previous work, we can reduce the size of our reachable belief set by removing beliefs within the same $\delta$-ball for some parameter $\delta_\ell$.

\subsection{Joint Belief Estimation} \label{sec:joint_belief_est}

\paragraph{Exact Reconstruction: Theoretical Possibility.}
When inferring other agents' beliefs through our pruning process, we generate both the posterior beliefs and the action-observation histories that led to them. Rather than using the beliefs directly, we can leverage these inferred actions and observations to update an estimated joint belief without approximation error. Using the estimates of the observations and actions, we can update the joint belief using:
\begin{align}
    \hat{\tilde{b}}^i_t (s^\prime) \propto O(\hat{\mathbf{o}} \mid \hat{\mathbf{a}}, s^\prime) \sum_{s} T(s^\prime \mid s, \hat{\mathbf{a}})\hat{\tilde{b}}^i_{t^-}(s)
\end{align}%
where $\hat{\tilde{b}}^i_{t^-}$ is the joint belief estimation from the previous time step, $\hat{\mathbf{o}} = (\hat{o}^{i, 1}, \ldots, o^i, \ldots, \hat{o}^{i,n})$, and $\hat{\mathbf{a}} = (\hat{a}^{i,1}, \ldots, a^i, \ldots, \hat{a}^{i,n})$.

While theoretically achievable, this exact approach presents computational challenges. The memory requirements increase substantially, as we must store $|\hat{\mathcal{B}}^{i,j}|$ beliefs for each agent plus $\prod_{j \neq i} |\hat{\mathcal{B}}^{i,j}|$ joint belief combinations, where the joint combinations can grow exponentially with the number of agents. The computational overhead similarly increases from $\sum_{j \neq i} |\hat{\mathcal{B}}^{i,j}|$ individual belief updates to $\sum_{j \neq i} |\hat{\mathcal{B}}^{i,j}| + \prod_{j \neq i} |\hat{\mathcal{B}}^{i,j}|$ total updates per time step. These computational challenges motivate consideration of approximate methods that capture the essential benefits while maintaining practical feasibility.

\paragraph{Practical Approximation: Conflation.}
After inferring beliefs of other agents, we can combine the inferred beliefs with the receiving agent's own belief to estimate a joint belief. Various methods exist for combining probability distributions \citep{genest_combining_dists_1986}. While mixture distributions are a straightforward approach, they require assigning and justifying potentially unequal weights.

An alternative method is conflation \citep{hill2011conflations}:
\begin{align}
\hat{\tilde{b}}^i(s) = \frac{b^i(s) \prod_{j\neq i}{\hat{b}^{i, j}(s)}}{\sum_{s^\prime \in \mathcal{S}} b^i(s^\prime) \prod_{j\neq i}{\hat{b}^{i, j}(s^\prime)}}.
\end{align}%

Unlike many combination methods, conflation possesses several desirable properties. Notably, conflation is not idempotent (i.e., $T(P, \ldots, P) \neq P$), which is beneficial when consolidating results from independent observations. As noted by \citet{hill2011conflations}, conflation minimizes the loss of Shannon information when consolidating multiple distributions, automatically prioritizes more confident beliefs, and requires no ad hoc weight assignments, making it a principled choice for belief combination. From a practical standpoint, conflation operates directly on available belief estimates without requiring the increased memory or computation of exact reconstruction, producing a single joint belief estimate that can be immediately used for action selection.

\subsection{Belief and Action Selection}
Using the suggested joint actions to prune the reachable beliefs and removing similar beliefs is effective in reducing the size of our estimated belief set. However, the belief subspace dominated by the suggested action can be composed of disjoint subsets, and pruning does not guarantee the reduction to a single belief

To form our set of estimated joint beliefs, we consider all possible estimated beliefs of other agents, resulting in at most $\prod_{j \neq i} |\hat{\mathcal{B}}^{i,j}|$ combinations. In practice, when the information implied by an action results in a small belief subspace, we often do not have many beliefs to consider. We demonstrate this in our experiments by sharing the alpha vector index instead of the action, thus sharing a single subspace region that is dominated by the optimal action. However, in cases where an action does not imply much information, the pruning is less effective, and we must determine how to use the set of estimated joint beliefs to select an action.

Rather than combining joint belief estimations (e.g., through centroids or averages), we propose a heuristic using counts for each unique belief. Counts increment when a similar belief is pruned, indicating how frequently each belief is reached through different paths. The estimated joint belief with the highest count is selected, breaking ties randomly to avoid bias. We then use this selected estimated joint belief to choose an action using a policy based on the assumption of shared observations and actions (a centralized joint policy).

This approach of maintaining belief counts and selecting based on weights balances computational efficiency and decision quality in our experiments, though effectiveness varies with problem characteristics. Future improvements could include optimal belief and action selection strategies for various problem scenarios like implementing history-based selection for more nuanced belief choice and using regret minimization across all estimated joint beliefs \citep{Auer2002} for an action selection strategy.

\subsection{Multiagent Control via Action Suggestions (MCAS) Algorithm} \label{sec:algo}

Our approach begins by solving $n+1$ MPOMDPs. For each agent $i \in {1, \ldots, n}$, we solve an MPOMDP where agent $i$ receives individual observations (observation space $\mathcal{O}^i$) but has control over all agents (action space $\mathcal{A}^1 \times \mathcal{A}^2 \times \cdots \times \mathcal{A}^n$). This results in policies $\pi^1, \ldots, \pi^n$ which have joint actions as the action space. We also require a policy $\tilde{\pi}$ that assumes joint observations and uses a joint belief, which can be generated by any suitable solver (online or offline).

The MCAS algorithm (\cref{alg:main_alg}) operates from the perspective of agent $1$, arbitrarily designated as the coordinating agent. This approach builds upon leader-based coordination but differs by integrating information from all agents. Unlike hierarchical action selection, it does not rely on a fixed communication order; instead, it treats all agents' suggestions equally to infer a comprehensive joint belief. The coordinating agent receives action suggestions from others, estimates a joint belief, and suggests a final joint action based on the centralized policy, which all agents then follow. This coordination role is not strictly necessary, but simplifies the discussion and presentation of our results. 

\begin{algorithm2e}[tb!]
\scriptsize
\caption{Multiagent Control via Action Suggestions} \label{alg:main_alg}
\KwData{
    $n$ \FlushRightComment{Number of agents}
    \newline $\mathcal{P}^1, \ldots, \mathcal{P}^n$ \FlushRightComment{Agents' MPOMDPs}
    \newline $\pi^1, \ldots, \pi^n$ \FlushRightComment{Agents' policies}
    \newline $\tilde{\mathcal{P}}, \tilde{\pi}$ \FlushRightComment{Joint MPOMDP and policy}
    \newline $\delta_\text{joint}, \delta_\text{single}$ \FlushRightComment{Similarity thresholds}
    \newline $\overline{B}_{\text{max}}$ \FlushRightComment{Maximum number of estimated beliefs}
}
\vspace{0.5em}
Initialize belief $b^1$ for agent 1 \\
Initialize surrogate belief sets $(\hat{\mathcal{B}}^{1,j}, w^{1,j}) = \{(b_0^j, 1.0)\}$ for $j = 2, \ldots, n$ \\
\While{not done}{
    Receive messages $\sigma^{j, 1}$ from agents $j = 2, \ldots, n$ \\
    \For{$j \gets 2$ \textbf{to} $n$}{
        $\hat{\mathcal{B}}^{1,j} \gets \textsc{PruneBeliefs}(\pi^j, \hat{\mathcal{B}}^{1,j}, \boldsymbol{\sigma^{j,1}}$) \\
        $\hat{\mathcal{B}}^{1,j} \gets \textsc{ReduceToMaxLimit}(\hat{\mathcal{B}}^{1,j}, \overline{B}_{\text{max}})$ \label{line:reduce}
    }
    $\hat{\tilde{b}} \gets \textsc{SelectJointBelief}(\{(\hat{\mathcal{B}}^{1,j}, w^{1,j})\}_{j=2}^n, b^1, \delta_\text{joint})$ \\
    $\tilde{\textbf{a}} \gets \tilde{\pi}(\hat{\tilde{b}})$ \\
    Broadcast $\tilde{\textbf{a}}$ to all agents \\
    Execute $\tilde{\textbf{a}}[1]$ and observe $o^1$ \Comment*[f]{Agent $1$'s action} \\
    $b^1 \gets \textsc{update}(\mathcal{P}^1, b^1, \tilde{\textbf{a}}, o^1)$ \\
    \For{$j \gets 2$ \textbf{to} $n$}{
        $\hat{\mathcal{B}}^{1,j}, w^{1,j} \gets \textsc{UpdateEstBeliefs}(j, \mathcal{P}^j, \hat{\mathcal{B}}^{1,j}, w^{1,j}, \tilde{\textbf{a}}, \delta_\text{single})$
    }
}
\vspace{0.5em}
\SetKwProg{Fn}{Function}{}{}
\Fn{\textsc{PruneBeliefs}($\pi, \hat{\mathcal{B}}, \sigma$)}{
    \Return $\{b \in \hat{\mathcal{B}} \mid \pi(b) = \sigma \}$
}

\end{algorithm2e}

While presented with a designated coordinator for simplicity, MCAS can operate in a fully decentralized manner once action suggestions have been shared. Each agent can independently maintain its estimated joint belief and select actions according to their MPOMDP policy, often achieving performance comparable to the coordinator-based method when effective pruning results in small belief sets. Alternative coordination strategies that leverage the shared action suggestions include random delays where agents execute the most recently received suggestion, belief set considerations where agents prioritize decisions from those with the smallest belief set size, and voting schemes for collective action determination.

The algorithm can be implemented using either actions or alpha vector indices as messages. When using alpha vector indices (which we call MCAS-$\alpha$), each agent communicates which alpha vector dominates their belief rather than just the resulting action. Since multiple alpha vectors can correspond to the same action but define distinct regions of the belief space, this provides a more precise subspace for pruning, leading to more effective belief state estimation.  While sharing alpha vector indices requires agents to have access to identical policies and greater computational coordination, MCAS-$\alpha$ demonstrates the effectiveness of our belief pruning approach when additional information is available.

\begin{algorithm2e}[t!]
\scriptsize
\caption{Update Estimated Beliefs} \label{alg:update}
\SetKwProg{Fn}{Function}{}{}
\Fn{\textsc{UpdateEstBeliefs}\label{line:updatesurrogate}($j, \mathcal{P}, \hat{\mathcal{B}}, w, \mathbf{a}, \delta_\text{single}$)}{
    $\hat{\mathcal{B}}_\text{new} \gets \emptyset$, $w_\text{new} \gets \emptyset$ \\
    \For{$i \gets 1$ \textbf{to} $|\hat{\mathcal{B}}|$}{
        \For{$o \in \mathcal{O}^j$}{
            $b^\prime \gets \textsc{update}(\mathcal{P}, \hat{\mathcal{B}}[i], \mathbf{a}, o)$ \\
            $w^\prime \gets w[i] + 1$ \FlushRightComment{Count-based approach} \\
            \If{$\forall b^{\prime\prime} \in \hat{\mathcal{B}}_\text{new}: \|b^\prime - b^{\prime\prime}\|_1 \geq \delta_\text{single}$}{
                $\hat{\mathcal{B}}_\text{new} \gets \hat{\mathcal{B}}_\text{new} \cup \{b^\prime\}$ \\
                $w_\text{new} \gets w_\text{new} \cup \{w^\prime\}$
            }
            \Else{
                $k \gets \text{argmin}_{b^{\prime\prime} \in \hat{\mathcal{B}}_\text{new}} \|b^\prime - b^{\prime\prime}\|_1$ \\
                $w_\text{new}[k] \gets w_\text{new}[k] + w^\prime$
            }
        }
    }
    \Return $\hat{\mathcal{B}}_\text{new}, w_\text{new}$
}
\end{algorithm2e}

\begin{algorithm2e}[b!]
\scriptsize
\caption{Select Joint Beliefs} \label{alg:select}
\SetKwProg{Fn}{Function}{}{}
\Fn{\textsc{SelectJointBelief}($\{(\hat{\mathcal{B}}^j, w^j)\}_{j=2}^n, b^1, \delta_\text{joint}$)}{
    $\mathcal{B}_\text{combined} \gets \emptyset$, $w_\text{combined} \gets \emptyset$ \\
    \For{$(\hat{b}^2, \ldots, \hat{b}^n) \in \hat{\mathcal{B}}^2 \times \cdots \times \hat{\mathcal{B}}^n$}{
        $b^c \gets \textsc{EstimateJointBelief}(b^1, \hat{b}^2, \ldots, \hat{b}^n)$ \label{line:combine} \\
        $w^c \gets \sum_{j=2}^n w^j[\text{index}(\hat{b}^j)]$ \FlushRightComment{Count-based approach} \\
        \If{$\forall b^\prime \in \mathcal{B}_\text{combined}: \|b^c - b^\prime\|_1 \geq \delta_\text{joint}$}{
            $\mathcal{B}_\text{combined} \gets \mathcal{B}_\text{combined} \cup \{b^c\}$ \\
            $w_\text{combined} \gets w_\text{combined} \cup \{w^c\}$
        }
        \Else{
            $k \gets \text{argmin}_{b^\prime \in \mathcal{B}_\text{combined}} \|b^c - b^\prime\|_1$ \\
            $w_\text{combined}[k] \gets w_\text{combined}[k] + w_c$
        }
    }
    $w_\text{normalized} \gets w_\text{combined} / || w_\text{combined} ||_1 $ \\
    $k \gets \argmax_i w_\text{normalized}[i]$ \\
    \Return $\mathcal{B}_\text{combined}[k]$
}
\end{algorithm2e}

The \textsc{EstimateJointBelief} function (\cref{line:combine} of \cref{alg:select}) can be implemented using various methods such as weighted averaging or conflation (\cref{sec:joint_belief_est}). If maintaining estimated joint beliefs from inferred observations, the \textsc{UpdateEstBeliefs} function (\cref{alg:update}) would need to return associated observations, and the belief combination process would involve updates for all possible observation combinations, potentially improving the accuracy of the joint belief estimate at the cost of increased computational complexity.

Pruning based on the suggested action is effective in practice; however, the number of reachable beliefs can still grow exponentially in the worst case. The \textsc{ReduceToMaxLimit} function (\cref{line:reduce}) limits the size of the belief set to $\overline{B}_\text{max}$. Our implementation computes the $\mathcal{L}_1$ norm between all belief pairs, sorts these distances, and iteratively removes the lower-weighted belief of the closest pair, adding its weight to the remaining belief, until reaching $\overline{B}_\text{max}$.

\section{Experiments} \label{sec:experiments}

All experiments were implemented and executed using Julia \citep{julia_bezanson2017} with the POMDPs.jl framework \citep{egorov2017pomdps}. Problem implementations were based primarily on originating papers, with additional references to the Multiagent Systems Planning Page \citep{masplan_start} and the Dec-POMDP page \citep{amato_decpomdp} to ensure consistency with previous work. For context, we include the best-reported results from Dec-POMDP solvers when available, noting that our approach's use of communication makes these comparisons informative but not equivalent.

\subsection{Benchmark Problems}
We tested MCAS on several Dec-POMDP benchmarks: Decentralized Tiger \citep{nair_tiger_2003}, Broadcast Channel \citep{Hansen_2004}, Meeting in a $2 \times 2$ Grid \citep{bernstein2005}, Meeting in a $3 \times 3$ Grid \citep{amato_incremental_2009}, Cooperative Box Pushing \citep{Seuken2007}, Wireless Networking \citep{pajarinen2011}, and Mars Rover \citep{amato_goals_2009}. For detailed problem descriptions and implementations, we refer readers to the original papers and our accompanying repository.

The original problems were designed without considering communication. In our experiments, we found that when we allowed one agent, using only its individual observations, to control all agents, it often achieved performance similar to a full MPOMDP (with shared observations and actions). To better demonstrate the value of integrating different beliefs, we introduced modifications to increase difficulty and emphasize the importance of different agent observations.

We use qualifiers to denote problem modifications from the original implementation in our results:
\begin{itemize}[itemsep=0pt]
    \item \textit{UI}: Uniform initial belief distribution.
    \item \textit{WP}: Added penalties for wall collisions or message sending.
    \item \textit{DP}: Modified Broadcast buffer fill probabilities for three agents ($0.2$, $0.4$, $0.4$).
    \item \textit{SS}: Meet $2 \times 2$, changed starting positions from corners to same row or column.
    \item \textit{AG}: Meet $3 \times 3$, rewarded agents for meeting at any grid location, not just two corners.
    \item \textit{SO}: Box Push with stochastic observations (\SI{50}{\percent} accuracy).
    \item \textit{$5$G}: Additional Mars Rover sampling site (accessible from the original top-right location).
    \item \textit{Meet $27$}: Expanded version of Meet $2 \times 2$ with $27$ grid locations. Observation space expanded to also include \textit{no walls} and \textit{both walls}.
\end{itemize}

\subsection{Baseline Methods and Implementation Details} \label{sec:implementation}

We compared MCAS against the following baselines:
\begin{itemize}
    \item \textit{MMDP}: Multiagent MDP assuming full observability.
    \item \textit{MPOMDP}: Multiagent POMDP with centralized control.
    \item \textit{MPOMDP-C}: MPOMDP policy with joint beliefs generated by conflating the true individual agent beliefs.
    \item \textit{MCAS$-\alpha$}: MCAS using alpha vector indices. Used conflation with similarity parameters $\delta_{\text{single}}$ and $\delta_{\text{joint}}$ set to \num{e-5}.
    \item \textit{MCAS}: As described in \cref{sec:algo}, using same parameters as MCAS$-\alpha$ with maximum estimated beliefs $\overline{B}_{\text{max}} = 200$.
    \item \textit{MPOMDP-I}: Single agent controls all agents, using only its individual observations.
    \item \textit{Independent}: Agents execute individual policies (assuming control of other agents), ignoring messages.
    \item \textit{Dec-POMDP}: Best reported results from literature (experiments not conducted by us).
\end{itemize}
Hyperparameter selection prioritized computational tractability: $\overline{B}_{\text{max}} = 200$ ensures reasonable simulation times while preserving belief diversity, and similarity thresholds $\delta_{\text{single}}, \delta_{\text{joint}} = 10^{-5}$ were set through empirical testing to merge functionally equivalent beliefs without impacting solution quality.

All POMDP policies were computed using SARSOP \citep{Kurniawati2008SARSOPEP}. Experiments for POMDP-based methods were conducted on a MacBook Pro with an Apple M1 Max processor and $32$ GB of memory, running each scenario $2000$ times. Results for these methods are reported with \SI{95}{\percent} confidence intervals. MMDP results represent the converged policy value and are reported without confidence intervals. Most problems used $50$ time steps with a discount factor of $0.9$, while the Wireless Network problem used $450$ steps and a $0.99$ discount factor.

\subsection{Results} \label{sec:results}

The results in \cref{tab:results} offer several insights into the performance of MCAS across various Dec-POMDP benchmarks. MPOMDP-C has similar performance to MPOMDP across all problems, suggesting that using conflation to combine beliefs is an effective approach, particularly in these scenarios where observations are independent. 

MCAS$-\alpha$ closely matches MPOMDP-C results through effective use of alpha vector indices for belief subspace refinement. While MCAS performs marginally worse using only shared actions for pruning, it still maintains effective joint belief estimates, achieving comparable results with less refined belief subspaces.

\begin{table*}[htb!]
\normalsize
\begin{center}
\ra{1.1}
\caption{Average cumulative discounted reward (with \SI{95}{\percent} confidence intervals) for various Dec-POMDP problems.}
\begin{adjustbox}{max width=\textwidth}
\begin{threeparttable}
    \begin{tabular}{@{}lrrrrrrrrrrr@{}}
        \toprule
        \multirow{2}{*}{\textit{Problem}}   & \multirow{2}{*}{\textit{Qualifiers}} &  \multirow{2}{*}{$\#$ \textit{Agents}} & \multicolumn{8}{c}{\textit{Solution Method}} \\
        \cmidrule{4-11}
                                            &            & & \textit{MMDP}  & \textit{MPOMDP}   & \textit{MPOMDP-C} & \textit{MCAS}$-\alpha$    & \textit{MCAS}      & \textit{MPOMDP-I} & \textit{Dec-POMDP}                                    & \textit{Independent}  \\
        \midrule
        \multirow{3}{*}{Dec-Tiger}          & ---           & 2 & $200.0$   & $59.5 \pm 0.9$    & $59.5 \pm 0.9$    & $58.5 \pm 0.9$            & $58.5 \pm 0.8$     & $34.3 \pm 1.7$    & $13.5$\tnote{[1]}\phantom{0}                      & $-68.1 \pm 3.5$       \\
                                            & ---           & 3 & $300.0$   & $108.5 \pm 1.0$   & $108.5 \pm 1.0$   & $108.5 \pm 1.0$           & $108.5 \pm 1.0$    & $82.1 \pm 1.5$    & ---                                                   & $-95.5 \pm 4.1$       \\
                                            & ---           & 4 & $400.0$   & $153.0 \pm 0.7$   & $153.0 \pm 0.7$   & $152.8 \pm 0.7$           & $152.8 \pm 0.7$    & $121.3 \pm 1.5$   & ---                                                   & $-121.4 \pm 4.4$      \\
        \grayrule           
        \multirow{2}{*}{Broadcast}          & ---           & 2 & $9.4$     & $9.4 \pm 0.0$     & $9.4 \pm 0.0$     & $9.4 \pm 0.0$             & $9.4 \pm 0.0$      & $9.4 \pm 0.0$     & $9.3$\tnote{[2]}\phantom{0}                      & $7.6 \pm 0.1$         \\
                                            & DP, WP        & 3 & $6.7$     & $6.6 \pm 0.0$     & $6.6 \pm 0.0$     & $6.6 \pm 0.0$             & $6.6 \pm 0.0$      & $5.5 \pm 0.0$     & ---                                                   & $-0.6 \pm 0.1$        \\
        \grayrule           
        \multirow{2}{*}{Meet $2\times 2$}   & ---           & 2 & $8.0$     & $6.4 \pm 0.1$     & $6.1 \pm 0.2$     & $6.1 \pm 0.2$             & $6.1 \pm 0.2$      & $5.9 \pm 0.1$     & $6.1$\tnote{*[3]}\phantom{0}                & $1.7 \pm 0.1$         \\
                                            & SS            & 2 & $8.4$     & $6.9 \pm 0.1$     & $6.8 \pm 0.1$     & $6.8 \pm 0.1$             & $6.8 \pm 0.1$      & $6.8 \pm 0.1$     & $7.0$\tnote{*[2]}\phantom{0}             & $2.3 \pm 0.1$         \\
                                            & UI, WP        & 2 & $8.7$     & $5.8 \pm 0.2$     & $5.3 \pm 0.2$     & $5.3 \pm 0.2$             & $5.3 \pm 0.2$      & $4.5 \pm 0.2$     & ---                                                   & $3.5 \pm 0.2$         \\
        \grayrule           
        \multirow{3}{*}{Meet $3\times 3$}   & ---           & 2 & $5.9$     & $5.8 \pm 0.1$     & $5.8 \pm 0.1$     & $5.8 \pm 0.1$             & $5.7 \pm 0.1$      & $3.6 \pm 0.1$     & $5.8$\tnote{[4]}\phantom{0}                      & $3.7 \pm 0.1$         \\
                                            & AG, UI, WP    & 2 & $8.1$     & $7.3 \pm 0.1$     & $7.3 \pm 0.1$     & $7.3 \pm 0.1$             & $7.1 \pm 0.1$      & $3.5 \pm 0.1$     & ---                                                   & $2.8 \pm 0.1$         \\
                                            & AG, UI, WP    & 3 & $7.2$     & $6.4 \pm 0.1$     & $6.4 \pm 0.1$     & $6.4 \pm 0.1$             & $6.2 \pm 0.1$      & $1.0 \pm 0.1$     & ---                                                   & $1.7 \pm 0.1$         \\
        \grayrule           
        Meet $27$                           & UI, WP        & 2 & $6.3$     & $2.2 \pm 0.1$     & $2.1 \pm 0.1$     & $2.0 \pm 0.1$             & $1.6 \pm 0.1$       & $0.6 \pm 0.1$     & ---                                                   & $0.6 \pm 0.1$         \\
        \grayrule           
        \multirow{2}{*}{Box Push}           & ---           & 2 & $240.1$   & $222.9 \pm 2.2$   & $223.4 \pm 2.1$   & $223.4 \pm 2.1$           & $223.0 \pm 2.2$    & $199.6 \pm 2.6$   & $224.4$\tnote{[4]}\phantom{0}                    & $163.6 \pm 3.4$       \\
                                            & SO            & 2 & $240.1$   & $204.3 \pm 2.5$   & $203.4 \pm 2.5$   & $203.2 \pm 2.5$           & $199.8 \pm 2.5$    & $178.8 \pm 2.7$   & ---                                                   & $138.5 \pm 3.8$       \\
        \grayrule           
        \multirow{2}{*}{Wireless}           & ---           & 2 & $-143.6$  & $-152.8 \pm 2.3$  & $-152.8 \pm 2.3$  & $-152.8 \pm 2.3$          & $-153.0 \pm 2.4$   & $-152.8 \pm 2.3$  & $-167.1$\tnote{$\dagger$[2]}\phantom{0}  & $-219.8 \pm 3.9$      \\
                                            & WP            & 2 & $-154.5$  & $-165.8 \pm 2.4$  & $-166.5 \pm 2.4$  & $-166.5 \pm 2.4$          & $-166.5 \pm 2.4$   & $-172.4 \pm 2.3$  & ---                                                   & $-240.2 \pm 4.1$      \\
        \grayrule           
        \multirow{4}{*}{Mars Rover}         & ---           & 2 & $29.2$    & $29.0 \pm 0.1$    & $29.0 \pm 0.1$    & $29.0 \pm 0.1$            & $29.0 \pm 0.1$     & $24.4 \pm 0.3$    & $26.9$\tnote{[4]}\phantom{0}                     & $26.0 \pm 0.2$        \\
                                            & UI            & 2 & $24.9$    & $23.9 \pm 0.1$    & $23.9 \pm 0.1$    & $23.9 \pm 0.1$            & $19.8 \pm 0.2$     & $16.4 \pm 0.2$    & ---                                                   & $15.3 \pm 0.2$        \\
                                            & UI            & 3 & $26.2$    & $25.2 \pm 0.1$    & $25.2 \pm 0.1$    & $25.2 \pm 0.1$            & $23.8 \pm 0.2$     & $19.7 \pm 0.1$    & ---                                                   & $16.6 \pm 0.1$        \\
                                            & $5$G, UI      & 2 & $21.4$    & $20.7 \pm 0.1$    & $20.7 \pm 0.1$    & $20.7 \pm 0.8$            & $18.0 \pm 0.2$     & $14.8 \pm 0.1$    & ---                                                   & $13.1 \pm 0.2$        \\
        \bottomrule
    \end{tabular}
    \begin{tablenotes}
        \scriptsize 
        \item[] \hspace{-0.5em} [1] \citet{Pajarinen_2011_best}; [2] \citet{MacDermed_2013_best}; [3] \citet{Amato_mealy_2010}; [4] \citet{Dibangoye_2014_best}
        \item[*] {The papers reporting the best scores for Meeting $2 \times 2$ do not discuss the initial state. We associated the best-reported result with an initial condition based on the MPOMDP solutions (which is an upper bound on Dec-POMDP results). Other reported scores: \citet{Pajarinen_2011_best}: $6.9$, \citet{amato_goals_2009}: $5.6$.} 
        \item[$\dagger$] {\citet{Dibangoye_2014_best} reported a value of $-140.4$, but we were unable to verify the implementations details. The reported value $-140.4$ is better than the performance of the MPOMDP on our implementation which implies there is a difference in implementation. Previously highest reported score prior to \citet{MacDermed_2013_best} was $-175.4$ by \citet{Pajarinen_2011_best}.}
    \end{tablenotes}
\end{threeparttable}
\label{tab:results}
\end{adjustbox}
\end{center}
\end{table*}

MCAS effectively pruned beliefs, with $|\hat{\mathcal{B}}^{1,j}|$ exceeding the maximum set size limit in only two problems: \SI{3.2}{\percent} of Meet $27$ and \SI{87.8}{\percent} of Box Push-SO runs. The largest performance decreases for MCAS compared to MCAS$-\alpha$ occurred in Meet $27$, Box Push-SO, Mars Rover-UI, and Mars Rover-$5$G-UI. This difference is due to MCAS$-\alpha$'s more effective pruning. \Cref{tab:results_max_belief} shows the maximum estimated belief set sizes for problems with a noticeable increase for MCAS. Despite larger set sizes, MCAS still achieved high performance approaching that of MCAS$-\alpha$. We anticipate this gap will decrease with improved belief selection.

Comparing MPOMDP and MPOMDP-I results reveals that in problems like Broadcast, Meeting, and Wireless, sharing observations provides no advantage over using only individual observations. While this suggests opportunities for simplification in certain multiagent problems, determining optimal leadership roles requires further study.

\begin{table}[htb!]
    \normalsize
    \begin{center}
    \ra{1.0}
    \caption{Maximum size of $\hat{\mathcal{B}}^{1,j}$ per simulation.}
    \begin{adjustbox}{max width=0.45\textwidth}
    \begin{threeparttable}
        \begin{tabular}{@{}lrrrr@{}}
            \toprule
            \multirow{2}{*}{\textit{Problem}}   & \multirow{2}{*}{\textit{Qualifiers}} &  \multirow{2}{*}{$\#$ \textit{Agents}} & \multicolumn{2}{c}{\textit{Solution Method}} \\
            \cmidrule{4-5}
                                                &            & & \textit{MCAS}$-\alpha$    & \textit{MCAS} \\
            \midrule
            Meet $3 \times 3$                   & ---      & 2 & $1.0 \pm 0.0$ & $2.5 \pm 0.0$      \\
            \grayrule           
            Meet $27$                           & UI, WP   & 2 & $1.5 \pm 0.0$ & $16.8 \pm 1.6$      \\
            \grayrule           
            Box Push                            & SO       & 2 & $4.8 \pm 0.1$ & $192.1 \pm 1.2$     \\
            \grayrule           
            Wireless                            & ---      & 2 & $1.0 \pm 0.0$ & $18.0 \pm 0.9$     \\
            \grayrule
            \multirow{2}{*}{Mars Rover}         & UI       & 2 & $1.0 \pm 0.0$ & $2.0 \pm 0.0$      \\
                                                & $5$G, UI & 2 & $1.0 \pm 0.0$ & $3.0 \pm 0.0$      \\
            \bottomrule
        \end{tabular}
    \end{threeparttable}
    \label{tab:results_max_belief}
    \end{adjustbox}
    \end{center}
\end{table}

A challenge in conducting these experiments was the generation of MPOMDP policies. While this process is more tractable compared to Dec-POMDP solvers, the complexity of solving MPOMDPs grows exponentially with the number of agents. The online execution of MCAS, however, did not pose a major computational burden, with all simulations conducted on a standard laptop. This balance between offline policy generation and lightweight online execution makes MCAS promising for practical multiagent problems.

\section{Conclusions and Future Work} \label{sec:conclusion}
This paper introduced the Multiagent Control via Action Suggestions (MCAS) algorithm for coordinating agents in partially observable environments. By leveraging suggested actions as a form of communication, MCAS demonstrated performance comparable to centralized methods across Dec-POMDP benchmarks, while maintaining computational efficiency. The algorithm effectively prunes the reachable belief space enabling belief inference of other agents which allows for the estimation of a joint belief.

Though the results of MCAS are promising, several research directions remain open. A key area is deeper theoretical analysis, including studying convergence properties of belief estimation, establishing performance bounds relative to centralized methods, and investigating information-theoretic properties of action-based communication. Another important area is relaxing strong assumptions, such as investigating how similar surrogate policies need to be to maintain performance and exploring cases where agents do not always follow suggestions. Extending MCAS to online solvers like AdaOPS \citep{wu_adaops_2021} and BetaZero \citep{moss2024betazero} would enable solving larger problems but requires efficient methods to estimate belief subspaces in real-time and handle stochastic online policies.

Our results indicate that action-based communication can be a powerful tool for multiagent coordination, potentially bridging the gap between decentralized and centralized approaches. This approach lays the groundwork for more intuitive coordination in human-agent teams, opening possibilities for mixed-initiative planning and decision making in real-world applications.

\bibliography{references}


\end{document}